\documentclass[5p,twocolumn,times]{elsarticle}
\usepackage{enumitem}
\usepackage{lipsum}
\usepackage{lineno,hyperref}
\modulolinenumbers[5]
\usepackage[pdftex]{color}
\usepackage[font=footnotesize,labelfont=bf]{caption}
\usepackage[font=footnotesize,labelfont=bf]{subcaption}
\journal{Journal of \LaTeX\ Templates}

\usepackage{amssymb}

\biboptions{compress}

\usepackage[figuresright]{rotating}

\begin{document}

\begin{frontmatter}

\title{Saltatory targeting strategy in rock-paper-scissors models}

\address[1]{Research Centre for Data Intelligence, Zuyd University of Applied Sciences, Paul Henri Spaaklaan 3D, 6229 EN, Maastricht, The Netherlands
}

\address[2]{School of Science and Technology, Federal University of Rio Grande do Norte, 59072-970, P.O. Box 1524, Natal, RN, Brazil
}

\address[5]{Quantum Industrial Innovation, National Industrial Training Service, Av. Orlando Gomes 1845, 41650-010, Salvador, Brazil}

\address[3]{Department of Computer Engineering and Automation, Federal University of Rio Grande do Norte, Av. Senador Salgado Filho 300, Natal, 59078-970, Brazil}

\address[4]{School of Mathematics, China University of Mining and Technology, Xuzhou 221116, China}

\author[1]{J. Menezes} 
\author[2,3,5]{R. Barbalho} 
\author[4]{Y. Sun}

\begin{abstract}
We explore how strategic leaps alter the classic rock–paper–scissors dynamics in spatially structured populations. In our model, individuals can expend energy reserves to jump toward regions with a high density of individuals of the species they dominate in the spatial game. This enables them to eliminate the target organisms and gain new territory, promoting species proliferation. Through stochastic, lattice-based simulations, we show that even when the energy allocated to jumping, as opposed to random walking, is low, there is a significant shift in the cyclic dominance balance. This arises from the increased likelihood of the leaping species successfully acquiring territory. Due to the cyclical nature of the game, the dominant species becomes the one that is superior to the jumping species.
We investigate how spatial patterns are affected and calculate the changes in characteristic length scales. Additionally, we quantify how saltatory targeting reshapes spatial correlations and drives shifts in population dominance. Finally, we estimate the coexistence probability and find evidence that this behavioural strategy may promote biodiversity among low-mobility organisms but jeopardise long-term coexistence in the case of high-mobility dispersal.
These results underscore the profound impact of novel foraging tactics on community structure and provide concrete parameters for ecologists seeking to incorporate behavioural innovation into ecosystem models.

\end{abstract}

\end{frontmatter}

\section{Introduction}
\label{sec1}

Space plays a fundamental role in shaping biodiversity, as demonstrated by extensive studies involving \textit{Escherichia coli} \cite{bacteria}. It is well established that three bacterial strains can coexist through cyclic dominance, following the rules of the classic rock-paper-scissors game: scissors cut paper, paper wraps rock, and rock crushes scissors \cite{mobilia2,Avelino-PRE-86-036112,uneven,sara}. Moreover, substantial evidence indicates that spatially structured ecosystems arise due to local interactions among organisms, which lead to the formation of distinct spatial domains \cite{Coli}. This phenomenon has also been observed in other natural systems such as Californian coral reef invertebrates and lizards \cite{coral,lizards}. Limited mobility promotes local interactions in these systems, fostering species coexistence and preserving biodiversity. Conversely, increased mobility tends to homogenise species distributions, often resulting in biodiversity loss \cite{Allelopathy}.

In addition, certain organisms have evolved non-standard movement strategies, such as directed jumps toward regions rich in prey or resources, rather than relying on simple random walks \cite{jump8,JOHNSTONBARRETT2025123049}. Jumping spiders prefer more active prey regardless of size, but evaluate prey size before attacking, approaching larger insects more cautiously. Consistent with optimal foraging theory, they reduce feeding time on less-preferred prey when better options are available \cite{jump1}. 

Plenty of evidence shows that saltatory dispersal or jump-based locomotion allows organisms to efficiently reach favourable environments while minimising exposure to predators or competitors. Planthopper insects use highly synchronized, rapid hind leg movements powered by thoracic muscles and a catapult mechanism to achieve exceptional jumping performance, reaching take-off speeds of $5.5$ $m/s$ and forcing over $700$ times their body mass. Their specialised hind limb anatomy, including fused coxae and a locking mechanism, supports this elite propulsion \cite{jump6}.
Also, it has been shown that parrotlets optimise energy use by coordinating leg and wing forces during foraging, minimising mechanical effort across distances and slopes. This strategy also sheds light on early flight evolution and offers insights for designing agile, climbing-flying robots \cite{jump2}.  

Elk migration in Colorado is driven by local, dynamic patterns of forage and snow, with significant variation among and within herds. Rather than following a single optimisation strategy, elk flexibly adjust migration timing and duration in response to changing environmental conditions \cite{jump3}.
Furthermore, Lévy walks are not a universally optimal search strategy, and their advantages diminish under more realistic conditions. Apparent power-law patterns in foraging data do not necessarily indicate actual Lévy behaviour or evolutionary adaptation \cite{jump4}.

This phenomenon has inspired Artificial Intelligence systems where
algorithms combining Lévy flight and artificial potential fields significantly improve multi-robot exploration efficiency in unpredictable environments \cite{jump5}. In addition, the natural phenomenon has inspired the creation of sophisticated algorithms in optimisation problems, enhancing
the traditional Fruit Fly Optimisation Algorithm by incorporating Lévy flight and an adaptive challenge mechanism to balance global exploration and local exploitation better. This significantly improved convergence accuracy and speed, as demonstrated on six benchmark optimisation problems \cite{jump7}.

There is ample evidence that many biological systems involve organisms adapting their behaviour in response to environmental cues \cite{ecology,Causes}. These behavioural strategies are crucial for individual survival, species persistence, and ecosystem stability \cite{MovementProfitable,Nature-bio}.
Within the context of the rock-paper-scissors model, behavioural movement strategies have been shown to affect species persistence and biodiversity maintenance significantly \cite{tenorio1,tenorio2,BARBALHO2024105229}. Defensive strategies—such as the Safeguard strategy \cite{Moura}, where individuals evade enemies, or Social Distancing during epidemic outbreaks \cite{combination,adaptivej,adaptivejj}—have proven effective in protecting individuals and enhancing the success of dominant species in spatial games \cite{Moura,adaptivej,adaptivejj}.

In this work, we investigate generalised rock-paper-scissors models in which individuals of one species employ a saltatory strategy inspired by ambush behaviour and adapted for efficient resource acquisition and competitive interaction. This strategy involves scanning the local environment and deciding to perform a directed jump based on the spatial distribution of competitors and resources.

We conduct our investigation using stochastic simulations based on the May-Leonard formulation of the rock-paper-scissors model. In this framework, organisms interact locally, and the total population size is not conserved \cite{Menezes_2023,tanimoto2,Szolnoki-JRSI-11-0735, Anti1,anti2,MENEZES2022101606,PhysRevE.97.032415}. Our study aims to adress the following questions:
\begin{enumerate}[label=(\roman*)]
\item What is the impact of saltatory targeting actions on spatial pattern formation?
\item How does intentional leap-based movement affect each species' typical sizes of spatial domains?
\item How does the frequency of jumps influence the long-term coexistence probability?
\end{enumerate}

This paper is organised as follows. In Section \ref{sec2}, we present the model and detail our simulation methodology. Section \ref{sec3} examines how the saltatory targeting strategy influences spatial patterns, and Section \ref{sec4} investigates the autocorrelation function and the characteristic length scales of these spatial domains. In Section \ref{sec5}, we explore how varying the jump frequency affects species coexistence, and we conclude with a discussion of the results in Section \ref{sec6}.

\begin{figure}
\centering	
\includegraphics[width=40mm]{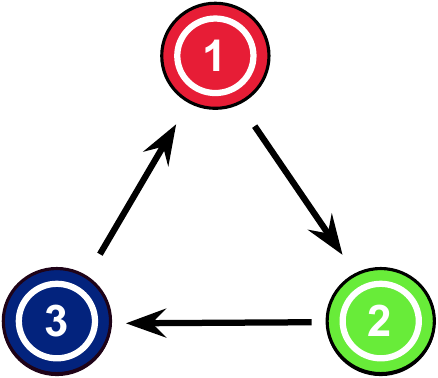}
\caption{Illustration of the rock-paper-scissors game rules. Selection interactions are represented by arrows indicating the dominance of organisms of species $i$ over individuals of species $i+1$.}
\label{fig1}
\end{figure}
\begin{figure*} 
\centering
    \begin{subfigure}{.19\textwidth}
        \centering
        \includegraphics[width=34mm]{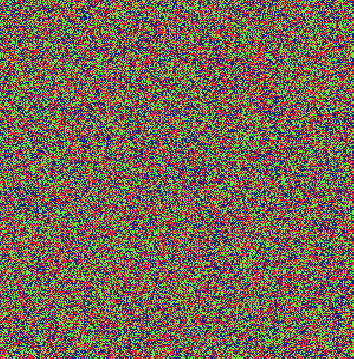}
        \caption{}\label{fig2a}
    \end{subfigure} %
   \begin{subfigure}{.19\textwidth}
        \centering
        \includegraphics[width=34mm]{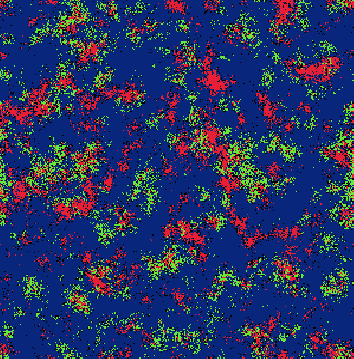}
        \caption{}\label{fig2b}
    \end{subfigure} 
            \begin{subfigure}{.19\textwidth}
        \centering
        \includegraphics[width=34mm]{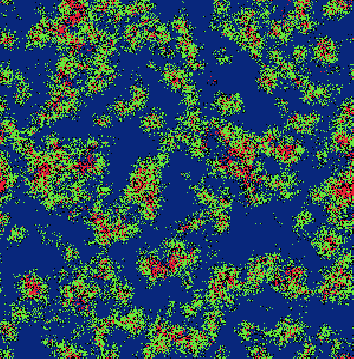}
        \caption{}\label{fig2c}
    \end{subfigure} 
           \begin{subfigure}{.19\textwidth}
        \centering
        \includegraphics[width=34mm]{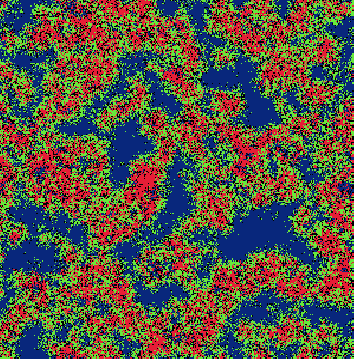}
        \caption{}\label{fig2d}
    \end{subfigure} 
   \begin{subfigure}{.19\textwidth}
        \centering
        \includegraphics[width=34mm]{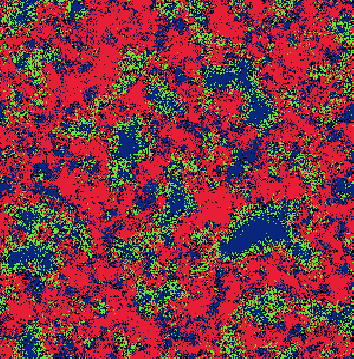}
        \caption{}\label{fig2e}
            \end{subfigure}
 \caption{Images captured during a simulation of the rock-paper-scissors model. Figure \ref{fig2a} shows the random initial conditions with species $1$ employing the saltatory strategy.
The lattice has $500^2$ grid sites, the timespan of $5000$ generations, $R=5$ and $\beta=0.1$.
The organisms' spatial organisation at $t=176$, $t=260$, $t=2360$, and $t=4760$,
 generations are showed in Figs.~\ref{fig2b} to ~\ref{fig2e}, respectively. The colours follow the scheme in Fig~\ref{fig1}; empty spaces appear as white dots. The dynamics of the organisms' spatial distribution is shown in video https://youtu.be/opR-EaHkZeQ.}
  \label{fig2}
\end{figure*}
\begin{figure}
   \centering
  \includegraphics[width=90mm]{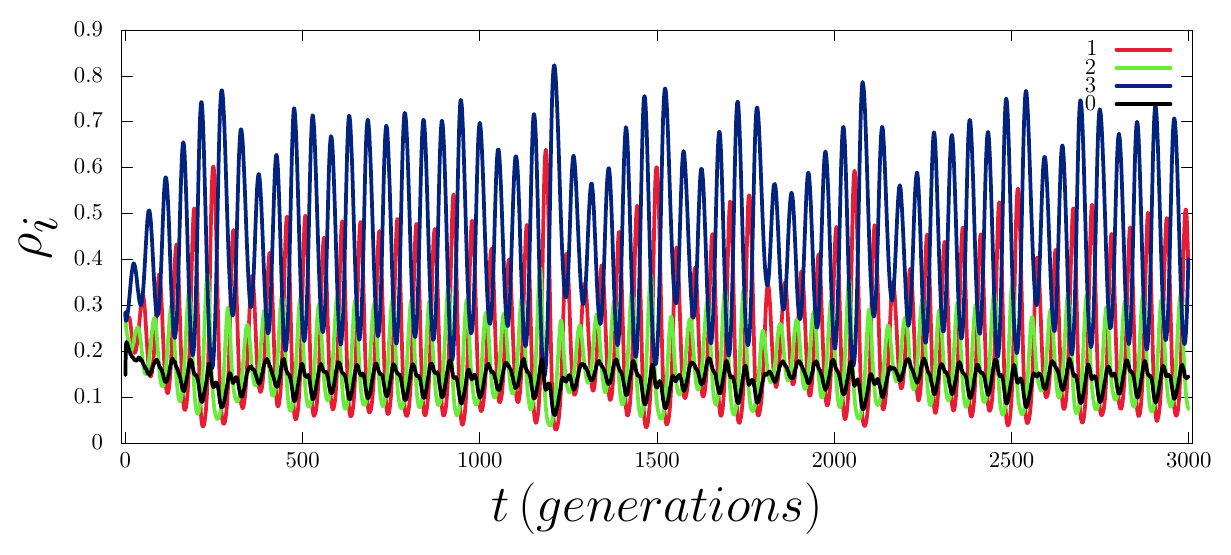}
  \caption{Temporal variation of species densities during the simulation in Fig.\ref{fig2}. Red, green, and blue lines show the population dynamics of species $1$, $2$, and $3$, respectively. The black line shows how the density of empty spaces changes with time.}
 \label{fig3}
\end{figure}
\begin{figure*}
	\centering
    \begin{subfigure}{.19\textwidth}
        \centering
        \includegraphics[width=34mm]{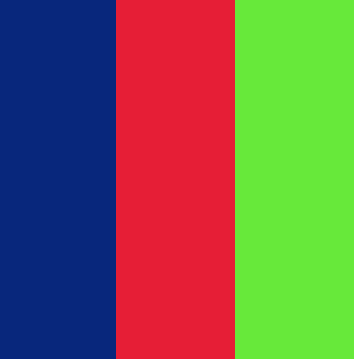}
        \caption{}\label{fig4a}
    \end{subfigure} %
   \begin{subfigure}{.19\textwidth}
        \centering
        \includegraphics[width=34mm]{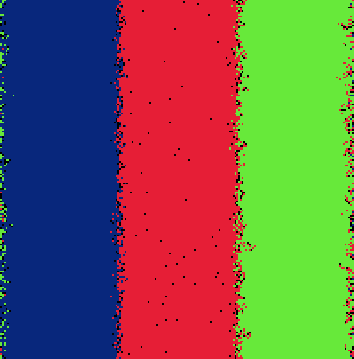}
        \caption{}\label{fig4b}
    \end{subfigure} 
            \begin{subfigure}{.19\textwidth}
        \centering
        \includegraphics[width=34mm]{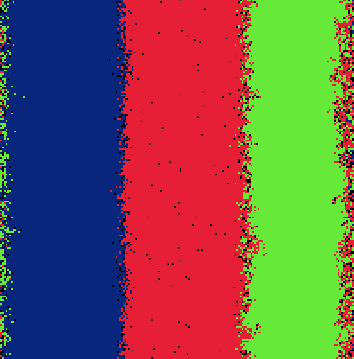}
        \caption{}\label{fig4c}
    \end{subfigure} 
           \begin{subfigure}{.19\textwidth}
        \centering
        \includegraphics[width=34mm]{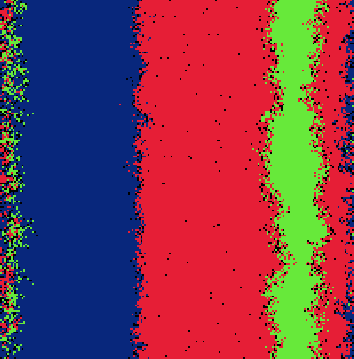}
        \caption{}\label{fig4d}
    \end{subfigure} 
   \begin{subfigure}{.19\textwidth}
        \centering
        \includegraphics[width=34mm]{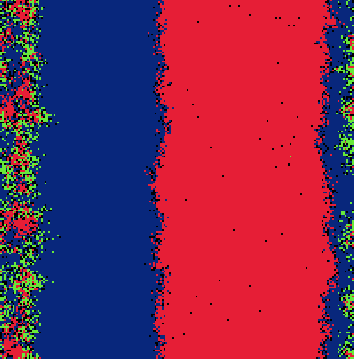}
        \caption{}\label{fig4e}
            \end{subfigure}\\
                \begin{subfigure}{.19\textwidth}
        \centering
        \includegraphics[width=34mm]{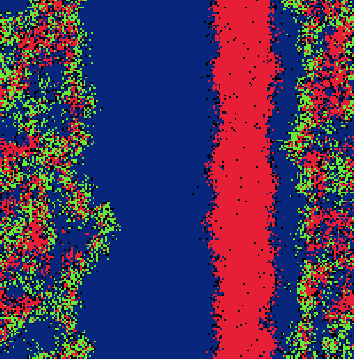}
        \caption{}\label{fig4f}
    \end{subfigure} %
   \begin{subfigure}{.19\textwidth}
        \centering
        \includegraphics[width=34mm]{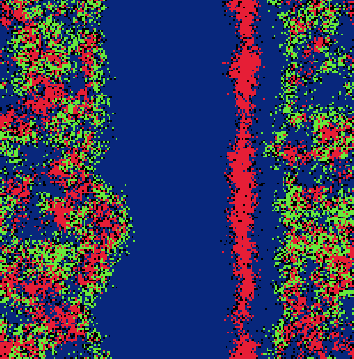}
        \caption{}\label{fig4g}
    \end{subfigure} 
            \begin{subfigure}{.19\textwidth}
        \centering
        \includegraphics[width=34mm]{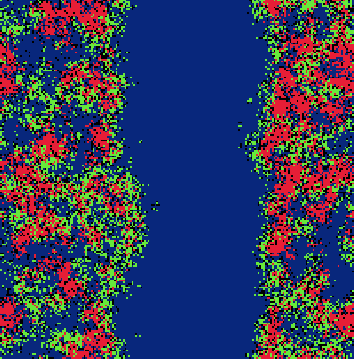}
        \caption{}\label{fig4h}
    \end{subfigure} 
           \begin{subfigure}{.19\textwidth}
        \centering
        \includegraphics[width=34mm]{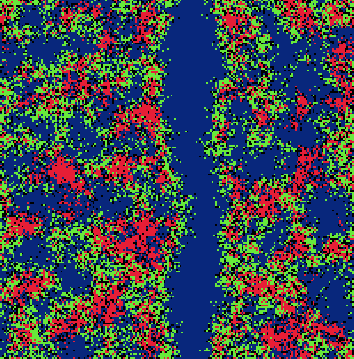}
        \caption{}\label{fig4i}
    \end{subfigure} 
   \begin{subfigure}{.19\textwidth}
        \centering
        \includegraphics[width=34mm]{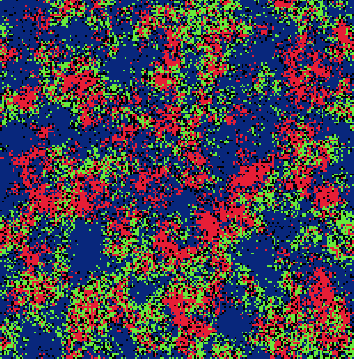}
        \caption{}\label{fig4j}
            \end{subfigure}
 \caption{Snapshots of a simulation starting from the prepared initial conditions shown in Fig.~\ref{fig4a}. Individuals of species $1$ (red) execute the jump attack strategy with $\eta=0.25$. The grid has
$300^2$ grid sites, and the timespan is $500$ generations.
Figures \ref{fig4b} to \ref{fig4j} show the spatial distributions of organisms at generations at generations $t=140$, $t=280$, $t=560$, $t=880$, 
$t=1140$, $t=176$, $t=260$, $t=2360$, and $t=4760$, respectively.
The colours follow the scheme in Fig~\ref{fig1}; empty spaces appear as white dots. The dynamics of the organisms' spatial distribution are shown in the video https://youtu.be/Mva-qmNl1gQ.}
  \label{fig4}
\end{figure*}
\section{Model and methods}
\label{sec2}
We investigate a three-species cyclic game, where species organisms compete for natural resources following the dynamics of the rock-paper-scissors model, as illustrated by Fig.\ref{fig1}. Accordingly, individuals of species $i$ eliminate organisms of species $i+1$, where $i$ ranges from $1$ to $3$; we express the cyclic identification as $i = i + 3\,\alpha$, with $\alpha$ standing for an arbitrary integer. 

In the standard model, all individuals of every species move randomly in the grid, giving a single step in one of the eight possible directions: moving means switching position with one of the immediate neighbours. In our model, individuals of species $i$ execute a behavioural strategy consisting of sporadically jumping to a distant place to be in a privileged position, surrounded by organisms of species $i+1$. This increases the chances of successful attack attempts, maximising their performance in the cyclic game. 

To implement the saltatory targeting strategy, we define the following parameters:
\begin{enumerate}
  \item Perception and jump range: each organism of species \(i\) continuously surveys its surroundings within a perception radius \(R\), which also defines the maximum distance it can relocate in a single jump. This enables the targeted detection of neighbouring individuals of species \(i+1\).  
  \item Jump energy allocation: whenever an organism is selected for movement, the saltatory targeting strategy
  can be executed with probability
  $\eta$, where $\eta$ is the saltatory energy fraction,
  a real number in the range $0\leq \eta \leq 1$. This factor represents the minimum proportion of the organism's energy reserves that must be allocated to jump.
  \item 
  Targeted relocation: if sufficient energy is available, the organism scans its perception zone to locate a potential empty site to jump. This empty space must be in a region where the local density of species \(i+1\) is high. It leaps there upon locating such a site, leaving its original grid site vacant. This action puts the individual in a perfect position to attack any organism of species \(i+1\) in the vicinity of the landing site.
  \item 
  Energy‐limited fallback: the search for potential empty sites to jump consumes an organism's energy. 
  We define $N_t$ as the total number of jump attempts made by individuals, which is an integer value less than the total number of points scanned.
  After this number of attempts, the saltatory targeting strategy is abandoned if the individual fails to find a high-density target site. Therefore, the organism performs a standard nearest‐neighbour random walk instead \cite{mobilia2}.
\end{enumerate}

The saltatory targeting strategy is adaptive: each individual of species $i$ autonomously identifies and leaps to the optimal location in real-time. All other species employ unbiased nearest‐neighbour random walks, as in standard spatial rock–paper–scissors models \cite{mobilia2}.

Simulations are performed on two‐dimensional square lattices of linear size \(N\) with periodic boundary conditions. Each of the \(\mathcal{N}=N^2\) sites can host at most one organism, so the system’s carrying capacity is \(\mathcal{N}\).

We adopt the May–Leonard framework for our simulations, in which the total population is not conserved \cite{leonard}. At \(t=0\), each of the three species is assigned the same number of individuals, which are distributed uniformly at random so that
\[
I_i \;\approx \;\frac{\mathcal{N}}{3}
\quad (i=1,2,3).
\]
The remaining sites are left initially empty.

Using the Moore neighbourhood, we consider that an organism can interact with one of its eight immediate neighbours. The possible actions follow the rules:
a) Selection: $i\ j \to i\ \otimes$, where $j = i+1$, and $\otimes$ means an empty space;
b) Reproduction: $i\ \otimes \to i\ i$;
c) Mobility: $i\ \odot \to \odot\ i$, with $\odot$ representing either an individual of any species or an empty site.

Interaction probabilities \(s\), \(r\), and \(m\) (for selection, reproduction, and mobility, respectively) are fixed and identical for all species. At each timestep, the algorithm proceeds as follows: i) select one active organism uniformly at random from the lattice; ii) sample an interaction type according to the probabilities \(s\), \(r\), and \(m\); iii) execute the chosen interaction:
    \begin{itemize}
      \item Selection or reproduction: choose one of the four nearest neighbours uniformly at random as the target site.
      \item Mobility: i) if the active organism is species $1$ and undertakes a saltatory targeting movement, determine the jump direction based on the local density of organisms of species $2$; ii) otherwise, select one of the four nearest neighbours uniformly at random and swap positions with it.
    \end{itemize}

The saltatory targeting strategy algorithm for species 1 proceeds as follows:
\begin{enumerate}[label=(\roman*)]
\item   
    A perception radius \(R\) is defined as the maximum range within which an individual can detect environmental cues. This is measured in lattice units.
  \item 
    A perception disc is constructed as the set of all lattice sites within the active individual's Euclidean distance \(R\).
  \item 
    A leap threshold is specified as a real parameter \(\beta \in [0,1]\). The leap‐trigger threshold corresponds to the minimum required local density of species $2$ around a landing site.
  \item 
  \(N_t\) attempts to find a site to jump. 
    \begin{enumerate}
      \item Randomly choose an empty site from the perception disc.  
      \item Check the empty space's eight neighbours; if all are occupied by species 2, the site is deemed suitable.
      \item If suitable, jump there (vacating the original site) and end the procedure.
      \item Otherwise, repeat (a)–(c) until either a suitable site is found or \(N_t\) trials have been exhausted.
    \end{enumerate}
  \item 
    If no suitable site is identified after \(N_t\) attempts, perform a standard nearest‐neighbour random walk: swap positions with one randomly selected adjacent site.
\end{enumerate}

All results presented in this article derive from simulations with interaction probabilities set to
$
s = r = m = 1/3$.
However, we have verified that our qualitative conclusions remain robust when using alternative probability sets.

\section{Spatial Patterns}
\label{sec3}

To initiate our analysis, we conducted a single simulation aimed at thoroughly examining the spatial pattern formation arising from the initial conditions in the rock-paper-scissors model with the saltatory targeting strategy. In this simulation, individuals of species $1$ employ the saltatory targeting strategy with a perception radius of $R=250$, saltatory energy fraction flight $\eta=0.25$, and leap threshold $\beta=1$ within a square lattice consisting of $500^2$ sites.

The spatial organisation of individuals at five distinct time steps: $t=0$, $t=140$, $t=280$, $t=560$, and $t=4760$, presented in Figures~\ref{fig2a}–\ref{fig2e}, respectively. Adhering to the colour scheme in Fig.\ref{fig1}, individuals of species $1$, $2$, and $3$ are represented in red, green, and dark blue, respectively, while unoccupied sites are indicated in black. The full temporal dynamics of the spatial configuration can be viewed in the video: https://youtu.be/opR-EaHkZeQ.

The simulation begins with randomly distributed individuals (Fig.\ref{fig2a}), leading to intense selection dynamics in the early stages. The introduction of saltatory targeting strategy by species $1$ induces significant asymmetry, resulting in a transient regime characterised by rapidly shifting territorial dominance among the three species. This phenomenon is observable in Figures~\ref{fig2b} to \ref{fig2e}, where emergent clusters of conspecifics are intermittently eroded by invading species.

In contrast to the classical rock-paper-scissors model, where spatial self-organisation commonly leads to stable spiral wave patterns, our modified system's spatial domains are highly unstable. For example, regions dominated by species~$2$ (green) are recurrently infiltrated by individuals of species~$1$ (red), a pattern mirrored across all species interactions. As a result, the system does not exhibit the coherent spiral structures typically associated with cyclic dominance. Instead, the network displays a persistent alternation in territorial control, devoid of long-term spatial regularity.

\subsection{Dynamics of species densities}

We also calculated the temporal dependence of the densities of organisms of each species during the entire simulation. The outcomes are shown in Fig.~\ref{fig3}, where the red, green, and blue lines show the dynamics of the density of individuals of species $1$, $2$, and $3$; the black line depicts the variation in the density of empty spaces. 

Figure \ref{fig3} shows that the temporal dependence of the species densities follows the symmetry inherent to the rock-paper-scissors game: alternating dominance with constant average density. However, there is a clear modification in the dynamics concerning the standard model, where all organisms of every species move randomly without jumping to long-distance regions.
Due to the saltatory targeting strategy of individuals of species $1$, their system becomes unbalanced, with a predominance of species $3$, with the highest average population throughout the simulation.

\subsection{Spatial pattern formation}

To better understand the effect of the saltatory targeting strategy in the individuals' spatial organisation, we performed a single simulation starting by the prepared initial conditions depicted in the snapshot of Fig.~\ref{fig2a}. Accordingly, each species occupies one-third of the grid - red, green and blue regions are filled by individuals of species 1, 2, and 3, respectively - there is no empty space at the beginning of the simulation. Because of the periodic boundary conditions, the only individuals that are in contact with organisms of a different species are located on the border of the rectangular domains.

In this simulation, individuals of species $1$ (red ones) execute the saltatory targeting strategy with a perception radius of $R=\mathcal(N)/2$, and the probability of flight attempt is $\eta=0.25$. The lattice contains $300^2$ grid sites, and the simulation timespan is $5000$ generations. 
Figures \ref{fig4a}, \ref{fig4b}, \ref{fig4c}, \ref{fig4d}, \ref{fig4e}, \ref{fig4f}, \ref{fig4g}, \ref{fig4h}, \ref{fig4i}, and \ref{fig4j} show the individuals' spatial organisation at: $t=0,5,10,25,45,100,120,140,190,220$, respectively. 
The dynamics of the spatial organisation during the whole simulation are available in the video https://youtu.be/Mva-qmNl1gQ.

When the simulation starts, the organisms on the border can interact using the selection rules. This means that individuals of species $i$ naturally kill individuals of $i+1$, thus creating empty sites, which are represented by black dots in Fig.~\ref{fig4b}. The appearance of empty spaces on the borders gives individuals of species $1$ the opportunity to execute the saltatory targeting strategy: some red organisms spring to the border where i) empty spaces are formed and ii) green organisms are presented. This means there are two borders to which they can leap: the border between blue and green regions and the border between red and green domains. However, there is a significant difference between the consequences of jumping to the different borders:
\begin{itemize}
\item 
Individuals of species $1$ that jump stochastically to the border between red and green areas benefit from the proximity of species $2$; thus, they are mainly successful in killing them and further reproducing and invading the region.
\item 
Although individuals of species $1$ that randomly leap to the border between blue and green take advantage of their position to kill individuals of species $2$, they place themselves close to their enemies of species $3$, significantly increasing the chances of being killed.
\end{itemize}
Therefore, the performance of the saltatory targeting strategy can both benefit and harm individuals, as we see in Figs.~\ref{fig4c} to ~\ref{fig4j}.


\begin{figure}
   \centering
  \includegraphics[width=90mm]{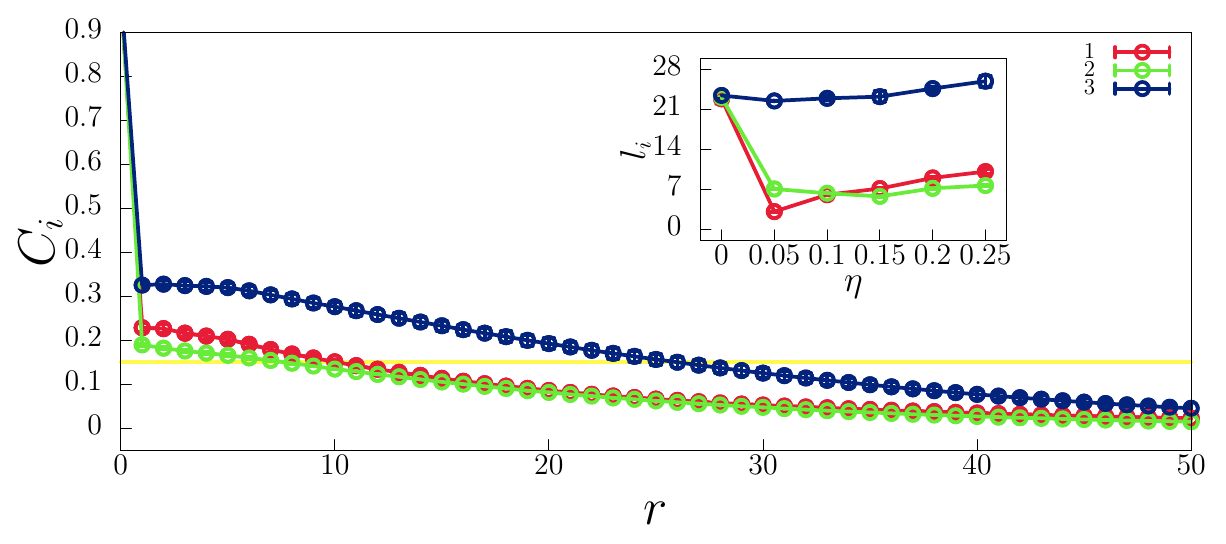}
  \caption{Autocorrelation function for each species in the rock-paper-scissors model with saltatory targeting strategy. We set the organisms' perception radius to $R = N/2$ and the flight frequency trigger to $\eta = 0.25$. The results were averaged over 100 simulations performed on a grid with $500^2$ sites, running for 5000 generations each. The colour scheme follows Fig.~\ref{fig1}, with error bars indicating the standard deviation. The inset panel shows the characteristic length scale for various values of $\eta$. The horizontal purple lines represent the threshold used to calculate the characteristic length scale displayed in the inset panel.}
 \label{fig5}
\end{figure}

\section{Spatial domains' characteristic length scale}
\label{sec4}

We now examine the impact of the saltatory targeting strategy on the scale of typical spatial domains for individuals within the same species. To commence this inquiry, we determine the spatial autocorrelation function, denoted as $C_i(r)$, in terms of the radial coordinate $r$ for individuals of each species, where $i=1,2,3$ \cite{combination}.

To specify the position $\vec{r}$ in the lattice occupied by individuals of species $i$, we introduce the function $\phi_i(\vec{r})$. Utilizing the mean value $\langle\phi_i\rangle$ for the Fourier transform:
\begin{equation}
\varphi_i(\vec{\kappa}) = \mathcal{F}\,\{\phi_i(\vec{r})-\langle\phi_i\rangle\}, 
\end{equation}
and the spectral densities 
\begin{equation}
S_i(\vec{\kappa}) = \sum_{\kappa_x, \kappa_y}\,\varphi_i(\vec{\kappa}).
\end{equation}

Next, we perform the normalised inverse Fourier transform to obtain the autocorrelation function for species $i$ as
\begin{equation}
C_i(\vec{r}') = \frac{\mathcal{F}^{-1}\{S_i(\vec{k})\}}{C(0)},
\end{equation}
which can be written as a function of $r$ as
\begin{equation}
C_i(r') = \sum_{|\vec{r}'|=x+y} \frac{C_i(\vec{r}')}{min\left[2N-(x+y+1), (x+y+1)\right]}.
\end{equation}
Determining the characteristic length scale for the spatial domains of species $i$, denoted as $l_i$ with $i=1,2,3$, involves setting the threshold $C_i(l_i)=0.15$.

Figure \ref{fig5} presents the results of a set of $100$ simulations for $R=\mathcal{N}/2$ and $\eta=0.25$. The mean value of the autocorrelation function is depicted using the colours in the scheme of Fig.~\ref{fig1}, and the error bars show the standard deviation. The yellow dashed line indicates the threshold used to compute the characteristic length scale $l_i$, defining the average size of the typical spatial domain occupied by each species. Our findings show that the saltatory targeting strategy performed by individuals of species $1$ creates an asymmetry in the spatial territory occupation, with individuals of species $3$ being more spatially correlated. The red and green dots also indicate that species $1$ and $2$ are the least spatially correlated.

The outcomes reveal that irrespective of the radius observed, the agglomeration of individuals of species~3 is larger. This means that the typical size of the spatial domain of species~3 is greater than that of the others, independent of the radius observed. Furthermore, the clustering of the organisms that perform the saltatory targeting dispersal, species~1, is bigger only for small scales, namely, for~$r \leq 10$; otherwise, the average size of regions fully occupied by individuals of species~1 or species~2 is approximately the same.

The inner figure depicts the dependence of the length scale~$l_i$ on $\eta$. The results show that, compared with the standard model, where no saltatory targeting strategy is considered ($l_i(\eta = 0) \approx 24$, for~$i=1,2,3$), the characteristic length scale for the flying species decreases, as well as for the species that is hunted. For a very low flight frequency, $\eta=0.05$, one has~$l_1 \approx 3$ and~$l_2 \approx 4$, while~$l_3 \approx 23$. This results from:
\begin{itemize}
    \item[(i)] even rare jumps by organisms of species~1 break the consistency of their own groups, facilitating the invasion of species~3 and subsequent group destruction;
    \item[(ii)] the landing of organisms of species~1 that leap into the interior of species~2 groups provokes further erosion of these groups.
\end{itemize}
Furthermore, we discover that, as saltatory movement become more frequent, the adverse effect on the formation of spatial patterns of species~2 becomes more significant since more groups are vulnerable to invasions by jumping organisms of species~1. Because of this, as~$\eta$ grows, the increase in~$l_1$ is more accentuated than that of~$l_2$. Namely, for~$\eta \gtrsim 0.1$, the spatial patterns of species~1 are larger than those of species~2.
\section{Coexistence Probability}
\label{sec5}

Finally, we investigate species coexistence by implementing random initial conditions running $1000$ simulations in lattices with $100^2$ grid points, for each value of \(m\) in the range (\(0.05 \leq m \leq 0.95\)) in intervals of $\Delta\,m=0.05$. The parameters were set as \(R = 50\), with the selection and reproduction rates defined as \(s = r = (1-m)/2\). 
For this experiment, the simulations ran for $10000$ generations. We then define coexistence as the presence of at least one individual of every species at the end of a simulation, while extinction occurs if any species is absent. Therefore, coexistence probability is calculated as the fraction of simulations resulting in coexistence. This analysis is repeated for several jump frequencies \(\eta\).

\begin{figure}[htbp]
  \centering
  \includegraphics[width=1.0\linewidth]{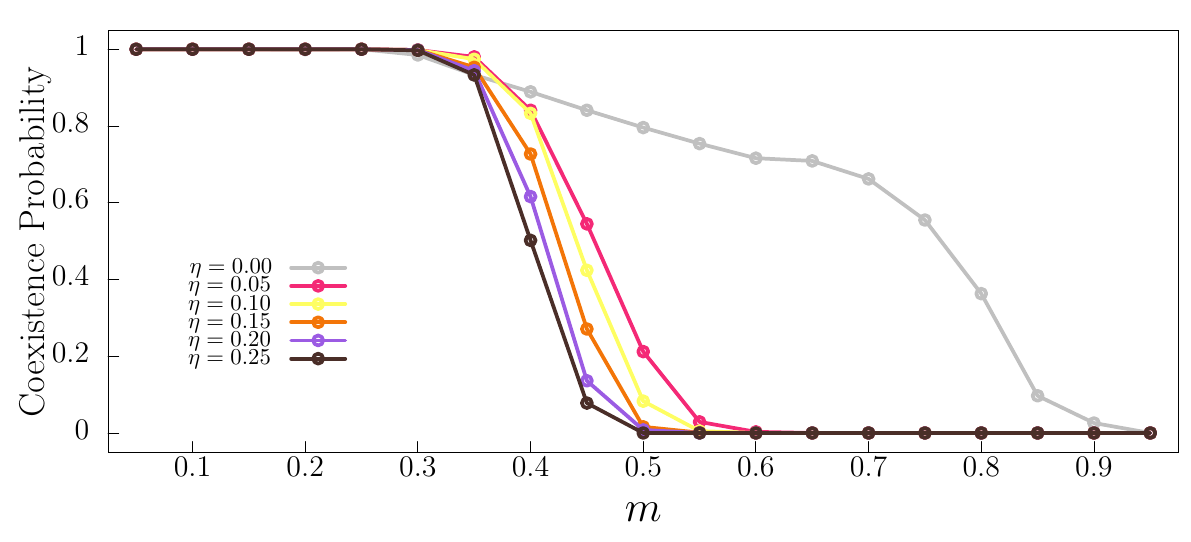} 
  \caption{Coexistence probability as a function of \(m\) for \(\eta = 0.05\) (red line), \(\eta = 0.10\) (yellow line), \(\eta = 0.15\) (orange line), \(\eta = 0.20\) (purple line), and \(\eta = 0.25\) (brown line). The standard model (\(\eta = 0.0\), no saltatory targeting strategy) is represented by the grey line.}
  \label{fig:11}
\end{figure}

Results indicate that biodiversity remains unaffected by saltatory energy fractions at very low mobility (\(m < 0.3\)), with coexistence probability at 100\%. Intriguingly, for a narrow intermediate range of \(m\) (\(0.3 \leq m \leq 0.35\)), saltatory targeting strategy enhances diversity, with coexistence probability increasing most significantly at lower \(\eta\). However, for \(m > 0.35\), saltatory targeting strategy jeopardises biodiversity: higher \(\eta\) correlates with sharper declines in coexistence probability. Beyond \(m > 0.6\), biodiversity collapses entirely, regardless of \(\eta\).

\section{Discussion and Conclusions}
\label{sec6}

In this work, we introduced and analysed a cyclic three-species system in which individuals of one species employ saltatory targeting dispersal. This long-range movement strategy dynamically relocates organisms into regions with a high density of individuals to beat in the spatial rock-paper-scissors game. Unlike stochastic Lévy flights previously applied to rock–paper–scissors frameworks \cite{jumprps} where relocation is random, our model endows individuals with environmental surveillance and decision-making autonomy. By allocating a fraction of metabolic energy toward strategic leaps, organisms of species $1$ detect favourable landing sites  — empty sites surrounded predominantly by species $i+1$ — and execute targeted jumps once an energy threshold is assumed. If energy reserves drop below this threshold, individuals revert to local, random diffusion, mirroring classical spatial RPS dynamics.

The mechanistic inclusion of an energy-dependent jump cost renders the model ecologically realistic and mathematically rich, bridging stochastic processes with nonlinear dynamics. Saltatory targeting can be interpreted as a nonlinear feedback mechanism: the more successful an organism is at conquering new territory, the greater its energy reserves and subsequent dispersal potential, introducing multiplicative coupling between local interactions and long-range mobility. Such feedback may generate spatial heterogeneity, influence correlation lengths, and reshape interface structures like pattern formation in reaction-diffusion systems.

Our study can be extended to simulate diverse behavioural strategies—such as defensive leaps avoiding enemy-rich zones \cite{BARBALHO2024105229,MENEZES2022101606}. By scanning for regions with minimal predator pressure, organisms perform protective jumps, analogous to safeguard strategies and collective defence phenomena observed in social species \cite{Moura}. This behaviour represents a natural extension of the safeguard strategy \cite{Moura} and could be further developed to resemble collective defence mechanisms observed in nature, such as those described in \cite{MENEZES2023104901}, where individuals aggregate with conspecifics to enhance protection within flocks. In this scenario, we expect that jumps aimed at forming defensive clusters would improve spatial organisation and promote species coexistence while reducing the organism's risk of death.

Moreover, the model can be generalised to more complex rock-paper-scissors systems involving a broader range of species interactions \cite{Avelino-PRE-89-042710,PhysRevE.99.052310,z2,ramylla}. In such frameworks, where competitive partnerships shape population dynamics, the interplay between short-range competition and long-range mobility—particularly when leap movements are involved—can significantly influence dominance patterns and interface structures. Sudden, targeted attacks by mobile individuals could alter the balance of power among alliances, introducing novel dynamics in the spatial distribution of species.

Our findings advance the theoretical understanding of movement ecology and spatial game theory and suggest experimental paradigms for empirical validation. In microbial consortia, engineered strains could be programmed to exhibit energy-dependent chemotactic leaps, offering a platform to test predictions on pattern emergence and biodiversity maintenance. Similarly, telemetry data from animal populations might reveal signatures of strategic long-range relocations in larger-scale ecosystems.
By coupling strategic, energy-based dispersal with local competitive interactions, our model synthesises concepts from nonlinear science, pattern formation, and behavioural ecology. It paves the way for future studies on the interplay between mobility strategies and ecological stability, with implications for conservation biology, epidemiology, and invasive species management.

\section*{Acknowledgments}
We thank CNPq/Brazil, ECT/UFRN, FAPERN/RN, ADSAI/Zuyd and Brightlands Smart Services Campus, and the Fundamental Research Funds for the Central Universities (Grant Nos. 2024KYJD2002 and 2023ZDYQ11005) for financial and technical support.
\bibliographystyle{elsarticle-num}
\bibliography{ref}

\end{document}